# Phase Coexistence of Mn Trimer Clusters and Antiferromagnetic Mn Islands on Ir(111)

Arturo Rodríguez-Sota[*], Vishesh Saxena, Jonas Spethmann, Roland Wiesendanger, Roberto Lo Conte[§], André Kubetzka, Kirsten von Bergmann[#]

*Institute for Nanostructure and Solid State Physics, University of Hamburg, 20355 Hamburg, Germany*

[*] arturo.rodriguez@physnet.uni-hamburg.de
[#] kirsten.von.bergmann@physnet.uni-hamburg.de
[§] Present address: Zernike Institute for Advanced Materials, University of Groningen, 9747 AG Groningen, The Netherlands

**Abstract:**

Clusters supported by solid substrates are prime candidates for heterogeneous catalysis and can be prepared in various ways. While mass-selected soft-landing methods are often used for the generation of monodisperse particles, self-assembly typically leads to a range of different cluster sizes. Here we show by scanning tunneling microscopy measurements that in the initial stages of growth Mn forms trimers on a close-packed hexagonal Ir surface, providing a route for self-organized monodisperse cluster formation on an isotropic metallic surface. For an increasing amount of Mn, first a phase with reconstructed monolayer islands is formed, until at full coverage a pseudomorphic Mn phase evolves which is the most densely packed one of the three different observed Mn phases on Ir(111). The magnetic state of both the reconstructed islands and the pseudomorphic film is found to be the prototypical antiferromagnetic Néel state with 120° spin rotation between all nearest neighbors in the hexagonal layer.

Material properties depend on composition and shape, and thus one way to tailor desired functionalities is by growth. Nano-scale particles can have radically different properties compared to bulk due to their increased surface and interface area over volume ratio, which can be exploited, e.g., for catalytic reactions. This is directly relevant for the research field of clusters, which ranges from chemical synthesis of particles for homogeneous catalysis to soft-landing or self-assembly of clusters on solid state substrates for heterogeneous catalysis [1], [2]. In general, once the optimum geometry and size has been identified, a monodisperse formation of clusters is desirable.

In the initial stages of metal-on-metal growth, single atoms diffuse over the surface until they hit other atoms, enabling the formation of small clusters that serve as nucleation centers. Depending on the diffusion coefficient, the deposition rate, and the thermal energy, additional incoming atoms either form new nucleation centers or attach to existing ones and thus contribute to growth [3], [4]. In contrast to film growth, where clusters are intermediate states, isolated clusters have also been prepared in superlattices, with characteristic distances between them. In those cases, either electronic or structural periodic modulations have been used as templates, such as electron standing waves on noble metal surfaces with surface states, e.g. Ag(111) [5], [6], or adsorbate superstructures made of an oxide [7], graphene [8], or hexagonal boron nitride [9]. Also charge transfer between clusters and the substrate has been identified as the origin of the formation of separated clusters due to a repulsive interaction between particles resulting from the electric dipole moments perpendicular to the surface [10].

In addition to self-assembly of clusters by chemical or physical methods, also a bottom-up approach of assembling clusters from individual atoms is possible by their manipulation on a surface with the tip of a scanning tunneling microscope (STM). In this respect, in particular magnetic atoms have been in the focus of attention, and the magnetic properties of clusters on surfaces have been studied as a function of size and geometry both experimentally and theoretically. Due to their high symmetry, in particular trimers on hexagonal surfaces have been investigated [11], [12], [13], [14]. Of special interest are trimers with antiferromagnetic coupling because the triangular structure gives rise to geometric frustration. The magnetic ground state of both an equilateral triangular trimer and a periodic hexagonal layer with nearest-neighbor antiferromagnetic exchange interaction is the Néel state with 120° between all neighboring spins [15], [16], [17], [18]. However, if other interactions come into play also a collinear antiferromagnetic state is possible, as was predicted for Mn trimers in equilateral triangular geometry on Au(111) and Cu(111) by first-principles [14], or found experimentally for a Mn monolayer on Re(0001) [19].

Here we report on an STM study showing that the deposition of Mn onto an Ir(111) surface leads to a monodisperse cluster phase of trimers in the low coverage regime. Beyond a critical coverage a phase coexistence of these trimers and reconstructed Mn monolayer islands is observed in the sub-monolayer regime. For coverages exceeding a fully closed monolayer another phase emerges, which is pseudomorphic with the highest density among the three different Mn phases. We study the magnetic properties of the Mn films and investigate the different types of clusters on the surface. A unidirectional switching of certain cluster types is possible. This switching shows a strong asymmetry in the critical threshold voltage, suggesting an impact of the electric field between tip and sample onto the switching mechanism. Careful analysis of the different cluster types enables a precise structure model. Finally, we suggest possible reasons for the Mn coverage dependent formation of these three phases with different Mn atom density, and the coexistence of two of them in the sub-monolayer coverage regime.

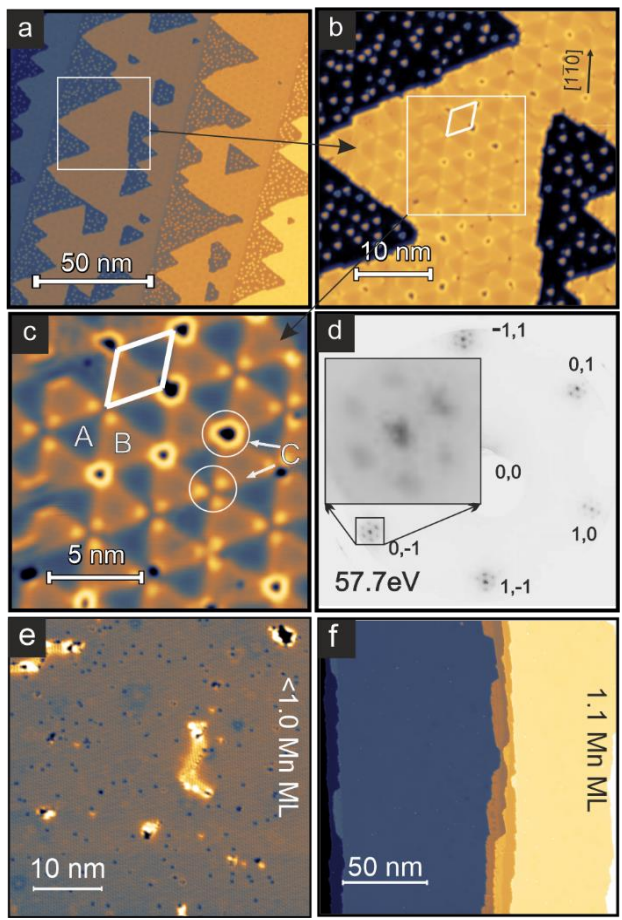

Figure 1. Growth of Mn on Ir(111). (a) Overview constant-current STM image of a sample of 0.7 atomic layers of Mn on Ir(111), partially differentiated for better visibility. (b) Reconstructed Mn monolayer islands and Mn clusters on Ir(111). (c) Closer view of the triangular reconstruction on the Mn monolayer island. (d) LEED pattern of a comparable sample. (e), (f) Constant-current STM images of Mn on Ir(111) with a coverage of nearly one atomic layer and just beyond one atomic layer, respectively. (Measurement parameters: a: $U$ = +100 mV; b,c,e: $U$ = +10 mV; f: $U$ = +30mV; all: $I$ = 1 nA, $T$ = 4 K).

**Experimental Results:**

Figures 1a,b show constant-current STM images of an Ir(111) single crystal surface partially covered with Mn where a coexistence of both Mn monolayer islands and clusters is observed (see methods for experimental details). At this coverage the islands have already coalesced, giving rise to triangular vacancy islands. This sample morphology, with both islands and clusters, occurs for sub-monolayer Mn coverage at all studied substrate temperatures (room temperature < $T$ < 200°C) during deposition (see Supplementary Fig. S1). Moreover, we find that the properties of the two phases, i.e., island phase and cluster phase, do not change, but only the ratio between their areas changes as a function of sub-monolayer coverage. In the following we will discuss the properties of the three different observed Mn phases one by one, starting with the Mn monolayer islands, then discussing the fully closed monolayer film, and finally turning to a detailed analysis of the cluster phase.

Mn monolayer islands exhibit a triangular reconstruction with a lattice constant of about 3.8 nm, i.e., about 14 nearest neighbor distances of the Ir substrate (see Figure 1b and the closer view in Figure 1c, where the unit cell of the reconstruction is indicated by a white diamond). The reconstruction is characterized by 3 distinct regions (cf. Figure 1c): darker triangles (A), brighter triangles (B), and the positions where six triangles meet (C). At these positions (C) we either observe a hole in the Mn layer or three bright dots. We attribute this superstructure to a lattice mismatch of Mn and Ir, which results in a Moiré-like structure, in which the A and B regions relate to Mn atoms roughly in the two different possible hollow sites, while the C positions indicate the on-top adsorption sites, which are apparently sometimes unoccupied leading to holes in the film (see also Supplementary Fig. S2). This superstructure gives rise to additional spots in the low energy electron diffraction (LEED) pattern, as seen in Figure 1d, with a ratio of roughly 1/15 between the reciprocal lattice vectors of the superstructure and the Ir(111). This is in good agreement with the size of the real space unit cell, indicating a lattice mismatch of around 7%.

When the amount of deposited Mn approaches a complete atomic layer the triangular reconstruction phase disappears, see Figures 1e,f, and the resulting Mn film is pseudomorphic, i.e., the Mn layer exhibits the same in-plane lattice constant as the underlying Ir(111) substrate. For coverages just below a complete layer, as in Figure 1e, only small remaining defects such as holes or areas reminiscent of the reconstruction are present. For coverages higher than one atomic layer a phase with a perfectly closed pseudomorphic Mn monolayer film is formed and small double layer Mn areas emerge, see Figure 1f.

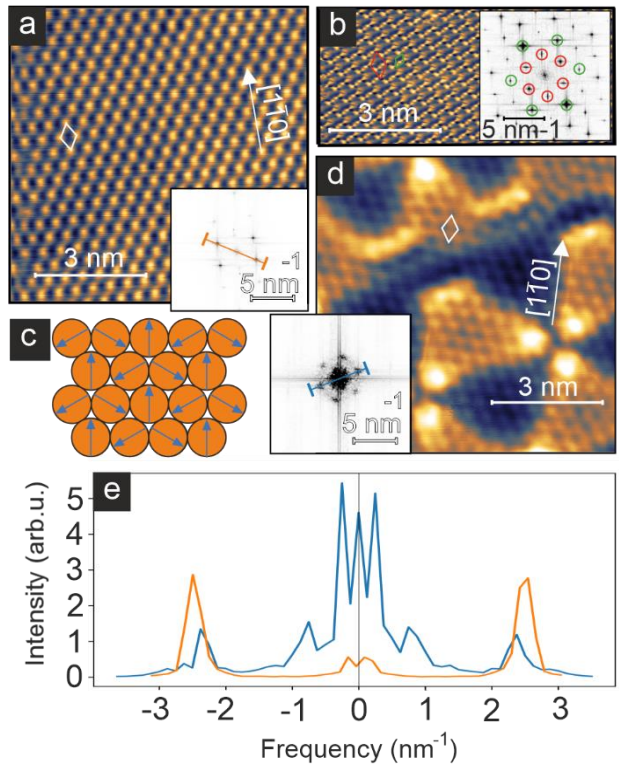

Figure 2. Magnetic ground state of the Mn monolayer. (a) Spin-resolved constant-current STM image of the pseudomorphic Mn monolayer on Ir(111); the hexagonal superstructure is of magnetic origin and characteristic for the Néel state; inset shows the FFT. (b) Magnetic atom manipulation image showing the atomic and the (√3x√3)R30° magnetic unit cell; inset shows the FFT, green circles mark the atomic periodicity whereas red circles mark the magnetic periodicity. (c) Spin model of the Néel state with 120° between nearest neighbor spins. (d) Spin-resolved constant-current STM image of the reconstructed Mn monolayer; the hexagonal superstructure indicated by the white diamond originates from the Néel state; inset shows the FFT. (e) Line profiles taken at the indicated positions in the FFT images of a,d. (Measurement parameters: a: $U$ = +3 mV, $I$ = 870 pA; b: $U$ = +10 mV, $I$ = 5 nA; d: $U$ = +5 mV, $I$ = 900 pA; all: $T$ = 4 K).

In order to investigate the magnetic properties of the Mn film a magnetic tip was used and spin-polarized (SP-)STM was employed, for which the spin-polarized contribution to the tunnel current scales with the projection of the local sample magnetization onto the tip magnetization direction (see methods) [20], [21], [22]. For the pseudomorphic Mn monolayer we find a hexagonal magnetic superstructure, see Figure 2a, which is rotated with respect to the atomic lattice and has a lattice constant of about 0.47 nm. The atomic-scale magnetic pattern typically appears very uniform, suggesting a commensurate spin structure. Indeed, images which also resolve the atomic lattice at the same time as the magnetic superstructure confirm this, see Figure 2b. The observed (√3 x √3)R30° magnetic superstructure is characteristic for the Néel state, with spins rotated by 120° between all nearest neighbors as shown in the sketch of Figure 2c [23]. This Néel state is the expected magnetic ground state for a hexagonal atomic lattice of spins with antiferromagnetic coupling between them.

When a magnetic tip is used for imaging, also the reconstructed Mn islands show a hexagonal pattern of similar size on top of their structural modulation, see Figure 2d. In this sample area hexagonal patterns of both bright dots and dark dots can be seen, which we interpret as two inversional domains of the Néel state. This is observed frequently for the reconstructed Mn, whereas the magnetic domains in the pseudomorphic Mn are typically larger. The fast Fourier transforms (FFT) of the STM images of Figure 2a and 2d are shown as insets, and line profiles across the principal spots of the respective Néel states at the indicated positions are displayed in Figure 2e. This direct comparison shows slightly (6%) larger reciprocal lattice vectors for the Néel state in the reconstructed Mn than in the pseudomorphic Mn film. This directly relates to the different atomic distances in the two cases and we conclude that the lattice mismatch of the reconstructed Mn compared to the substrate is about 6%, in reasonable agreement with the size of the reconstruction in real and reciprocal space as derived from the data shown in Figure 1. In addition the direct comparison of the magnetic unit cell sizes of the two Mn phases provides the information that the reconstructed Mn monolayer is expanded with respect to the Ir(111) surface, meaning that the Mn atom density in the reconstructed phase is only about 87% of that in the pseudomorphic phase. This is quite surprising, as 3$d$ films on 5$d$ substrates typically incorporate more atoms into the film, i.e., reconstructed layers usually have a higher atom density compared to the substrate [24].

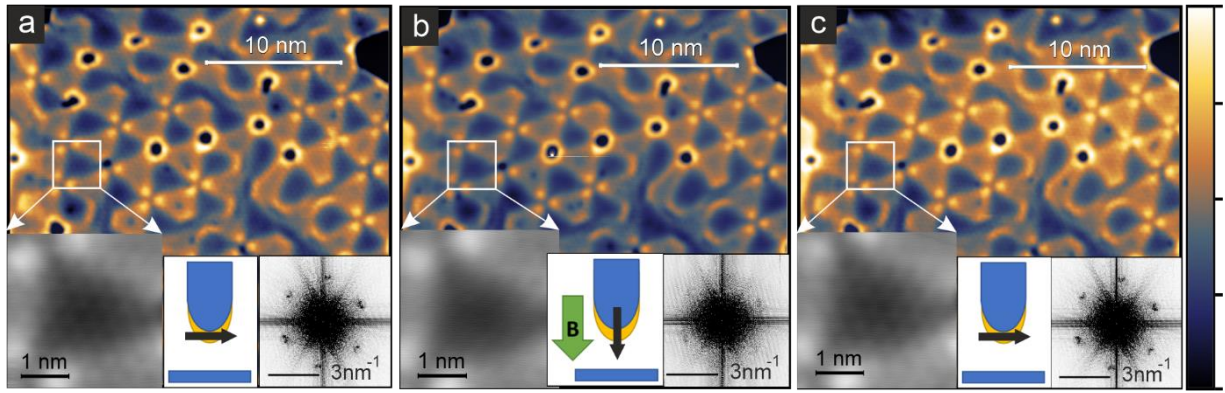

Figure 3. Néel state in the reconstructed Mn monolayer. (a),(b),(c) Spin-polarized STM measurements of the same sample area with different tip magnetization directions as indicated in the sketches. Insets show higher resolution data of the indicated areas with a $\Delta z$ of 40 pm. The FFTs have been obtained from the entire square overview images. (Measurement parameters: a: $B$= 0 T; b: $B$ = -2 T; c: $B$= 0 T; all: $U$ = +9 mV, $I$ = 1.5 nA, $T$ = 8 K; Fe-coated W tip which is magnetized in the surface plane without external magnetic field but aligns with an applied field as indicated in the sketches).

Figure 3 shows a measurement series with an Fe-coated W tip, that is magnetized parallel to the surface at zero magnetic field but can be aligned perpendicular to the Mn film by an applied magnetic field (see sketches). The reconstruction of the Mn monolayer is clearly seen. The enlarged view in Figure 1a shows the characteristic Néel superstructure, and the FFT of the overview image also shows the six spots at the positions expected for the magnetic Néel state. This magnetic signal at zero magnetic field originates from the in-plane sample magnetization components. When an out-of-plane magnetic field of -2T is applied, see Figure 3b, the tip magnetization aligns with the external field, see sketch, whereas the sample magnetization remains unaffected due to the compensated magnetization of the antiferromagnetic structure. The FFT of this SP-STM measurement does not show the Néel structure related spots. This demonstrates that the magnetic contribution to the tunnel current vanishes for this tip implying that the Néel state is magnetized fully in the surface plane. The original magnetic contrast of Figure 3a is recovered after switching off the out-of-plane magnetic field, see Figure 3c, with reappearing magnetic spots in the FFT due to the in-plane sensitivity of the magnetic tip. This measurement series shows that there are no out-of-plane sample magnetization components, and instead the magnetic moments in the Néel state are fully in the plane of the film.

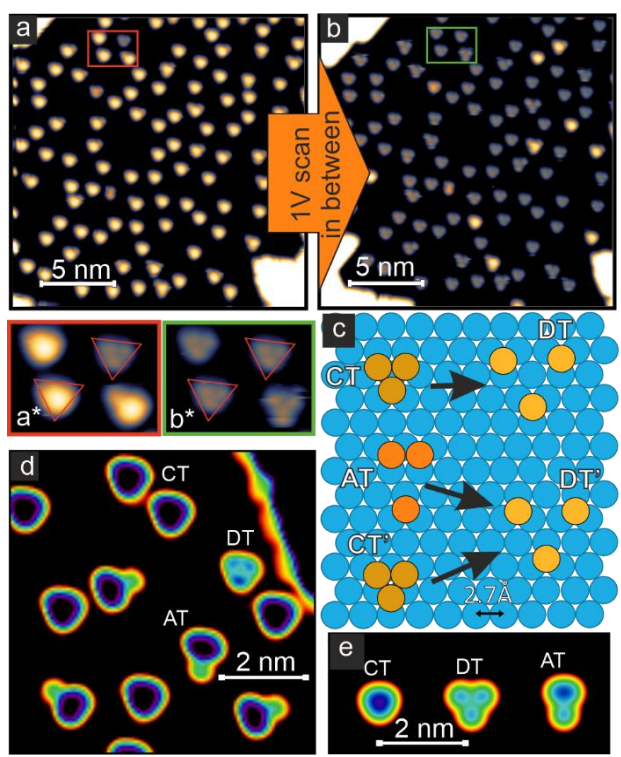

Figure 4. Cluster analysis. (a),(b) Constant-current STM images of the cluster phase before and after scanning the area with a bias voltage of +1 V, respectively; the enlarged views show four clusters. (c) Structure model of the different cluster types. (d) Different clusters, enlarged view of the red area indicated by the box in Figure 1b. (e) STM constant-height simulations of a compact trimer (CT), a dilute trimer (DT), and an asymmetric trimer (AT) on a hexagonal surface. (Measurement parameters: a,b: $U$ = +50 mV, $I$ = 100 pA, $T$ = 80 K; d: $U$ = +10 mV, $I$ = 1 nA, $T$ = 4 K.)

In the following, we concentrate on the Mn cluster phase, which is found to coexist with the reconstructed Mn islands, i.e. the clusters are adsorbed directly on the Ir surface, see Figure 4a. In this coexistence regime we find a cluster density of around 0.08 – 0.24 cluster/nm$^2$, with no clear correlation of the exact value to the sample area covered by the reconstructed Mn phase or the substrate temperature during Mn growth (see Supplementary Fig. S3). The clusters are statistically distributed in the areas between Mn islands with cluster centers separated by at least 1.3 nm, i.e, roughly 5 atomic distances of the Ir substrate (see Supplementary Fig. S3). If they would form a perfectly hexagonally ordered phase with a density of 0.15 cluster/nm$^2$, their distance would be 2.8 nm. If the minimal distance of 1.3 nm would be realized in a perfect hexagonally ordered cluster phase its density would amount to 0.6 cluster/nm$^2$.

A close inspection of Figure 4a shows that different cluster types can be identified, which can be discriminated by their height and symmetry. Typical cluster heights are 80% and 50% of the Mn island height, generating a distinction between what we call 'high clusters' and 'low clusters'. The ratio between the different cluster types varies depending on the sample preparation. Most of the high clusters have three-fold symmetry, however, some have an asymmetric shape and can occur in one out of three rotations (see also Figure 4d). The low clusters typically have three-fold symmetry with three distinct maxima around a local minimum in the center of the cluster, see enlarged view of Figure 4a. The heights of the maxima are on the order of the one typical for single atoms, suggesting that these are three single Mn atoms, i.e., a Mn trimer. The relative orientation of the maxima is the same as the atomic structure of the surface and their distance is around 0.4 nm. The nearest neighbor distance of the substrate is 0.27 nm, however, because of the large apparent diameter of single atoms we anticipate that their real spacing is significantly larger than the measured one and postulate that the low clusters are made out of three Mn atoms spaced with two Ir atomic distances.

The high clusters can be changed irreversibly by scanning the sample at higher bias voltage, e.g., at +1 V, which turns nearly all of them into low clusters in the scanned sample area (compare Figures 4a and 4b; see Supplementary Fig. S4 for complete dataset). The closer view image shows that the two high clusters on the left have switched to become the previously identified low Mn trimers. This suggests that also the high clusters are Mn trimers, albeit with a different geometry. We propose that the high clusters are close-packed trimers (CT), whereas the low trimers can be referred to as dilute trimers (DT). Given that the center of mass is conserved during the switching, see triangular marks in the insets, we arrive at the structure model shown in Figure 4c, where a surface Ir atom is located below the center of the high and low trimers CT and DT. Also the asymmetric high clusters switch, see bottom right cluster of the enlarged views of Figures 4a,b, and while the resulting low cluster looks very similar to the DT, it is more unstable and a hopping of atoms is observed during imaging. With this information we suggest cluster structures for asymmetric trimers (AT) and unstable dilute trimers (DT') as shown in Figure 4c. We find that also some compact trimers transition into DT', and conclude that they have the same geometry as the CT but a different relation to the substrate and call them CT'. A switching in the reverse direction, i.e., from low clusters to high clusters, has not been observed, and we conclude that at the measurement temperature of $T$ = 80 K the high clusters are metastable. Looking at the more unstable DT's at $T$ = 80 K (see Figure 4b inset and Supplementary Fig. S4), it is interesting to note that, while atoms can hop within the cluster, the trimer is still one entity despite this large Mn-Mn distance.

We have simulated STM images for these different cluster geometries on a hexagonal surface assuming atomic *s* orbitals [25], see Figure 4e, and the result agrees nicely with our experimental data, see Figure 4d. Very rarely we also find other cluster types, e.g. dimers, but nearly all of the clusters observed belong to one of the trimer types shown in the structure model of Figure 4c, demonstrating that in this system a monodisperse cluster formation by self-assembly is achieved. This is also in agreement with atom manipulation experiments where clusters were taken apart always yielding 3 atoms, and reassembling them into different cluster types (see Supplementary Fig. S5).

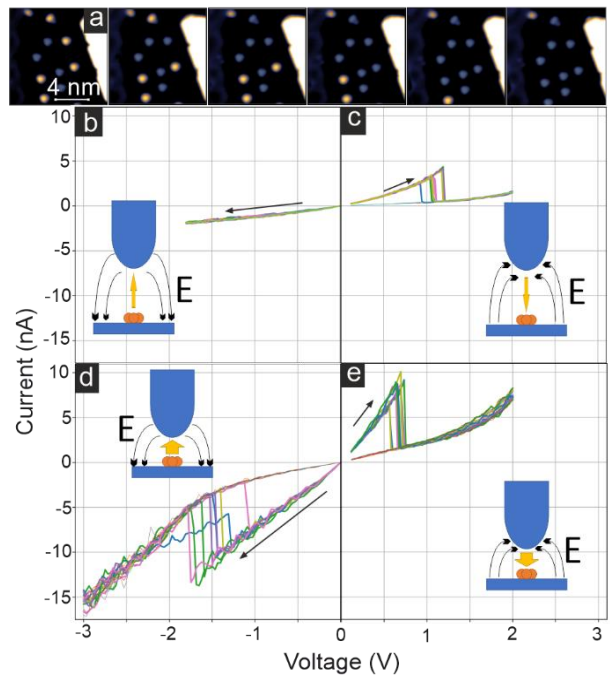

Figure 5. Switching of clusters. (a) Exemplary series where clusters are switched one by one from the high to low cluster type with local tunnel currents. (b), (c), (d), (e) Series of bias voltage sweeps from small to large |$U$| performed over different high clusters with constant tip-sample distance; the drop in |$I$| indicates a switching to low clusters; b,c larger and d,e smaller tip-cluster distance. (Measurement parameters: a: $U$ = +10 mV, $I$ = 1 nA; stabilization parameters before bias voltage sweeps: b-e: $U$ = +110 mV; b,c: $I$ = 200 pA; d,e: $I$ = 1 nA; all: $T$ = 4 K).

To obtain additional information about the conditions for the switching of high clusters to low cluster we positioned the tip over different high clusters, and performed bias voltage sweeps with constant tip-sample distance. In this way, clusters can be switched one-by-one, see image series in Figure 5a. The measured current responses for several individual compact trimers during the bias voltage sweeps are shown in Figure 5b-e for sweeps both in the positive and the negative bias voltage direction, and at larger and smaller tip-sample distances. A sudden drop in the obtained current indicates that the cluster changes its configuration, i.e., from a compact trimer to a dilute trimer (CT to DT). For comparably large tip-sample distance and negative bias voltages up to -1.8 V we do not observe a switching of high clusters to low clusters (Figure 5b); in contrast, we find that towards positive bias voltages (Figure 5c) the switching occurs repeatedly at about +1.1 V. When the tip-sample distance is decreased, this threshold voltage for switching with positive bias decreases to +0.7 V, see Figure 5e. The currents at which the switching is detected are about 3.5 A and 8.4 A for larger and smaller tip-sample distance, respectively, ruling out a mere current-induced effect. The related average switching powers are roughly 3.9 nW and 5.9 nW, respectively. With a smaller tip-sample distance also a switching with negative voltage sweeps is possible with an average threshold voltage of -1.45 V (11.0 nA and -16 nW), see Figure 5d. This strong dependence of the threshold voltage for switching on the bias polarity suggests that the electric field between tip and sample is involved in the switching mechanism; this could occur for instance in the case that the clusters exhibit an electric dipole moment, as was reported for Eu clusters on graphene on Ir(111) [10]. An electric field between tip and sample would enforce or counteract an electric cluster dipole moment, leading to a stabilization or destabilization of one cluster configuration with respect to another one with a different electric dipole moment. Such a scenario could explain the asymmetry of the threshold voltage; the possibility to switch the cluster with both polarities, however, also requires an additional mechanism, which might be related to hot electrons and thus depend on current and voltage.

**Discussion:**

We would now like to elaborate on possible explanations for the three different Mn phases on Ir(111) that occur for a coverage up to one complete monolayer. They exhibit a different density of Mn atoms: in relation to the pseudomorphic layer we find that the reconstructed Mn islands have a density of only 87%, and in the cluster phase the density is about 3-4% (albeit not dispersed as single atoms but as trimers). We will first look at coverages close to a monolayer, i.e., the transition from the reconstructed island phase to the pseudomorphic phase. When the amount of Mn is about 87% of the number of surface Ir atoms, the reconstructed Mn would cover the entire surface. Further deposition can then lead to either the formation of Mn double layer islands on top of the reconstructed monolayer, or to an incorporation of Mn atoms into the expanded reconstruction, thereby increasing the atom density within the Mn monolayer. The latter scenario is observed in our case, and must be related to the balance of free surface and interface energies. These parameters are not known, however, coverage-dependent phases with different densities have been observed before, for instance for nanowires of rare earth elements on W(110) [26].

When the coverage is lower, we observe the coexistence of reconstructed Mn islands and the cluster phase. As the coverage is increased the reconstructed islands grow at the expense of the cluster phase, which maintains roughly the same density. Two scenarios are possible: either the reconstructed Mn islands swallow the Mn clusters as they expand, or the clusters travel to the Mn islands and attach themselves. A close look at the perimeter of the Mn islands shows that typically straight edges and complete triangles of the Moiré-like reconstruction are found at the island rim. One of these triangles contains on the order of 120 atoms, i.e., the equivalent of about 40 trimers. This suggests, that island growth proceeds in steps, in which an appropriate number of Mn atoms equivalent to one or more of these triangles is incorporated at the island edge at once. This process would lead to an instantaneous drop in the local cluster density, which can explain the variation that we observe for this parameter (Supplementary Fig. S3). This can be understood in terms of a supersaturation, where the addition of Mn atoms during the growth process continuously generates trimers on the surface. Once a critical density in the cluster phase is reached, Mn is removed from the cluster phase and added to the reconstructed islands. This step-wise transfer of Mn atoms from the cluster phase to the island phase stops either when the Mn deposition is interrupted, or when the reconstructed Mn covers the entire surface.

A precondition for this scenario is that the trimers must be mobile at the temperature at which the growth happens, in our case at or above room temperature. STM measurements of samples with sub-monolayer coverage of Mn on Ir(111) at room temperature show the reconstructed islands, however, the area between the islands typically looks empty or fuzzy (see Supplementary Fig. S6). At intermediate temperatures sometimes agglomerates of clusters are observed, whereas at or below $T$ = 80 K the trimers remain isolated and can be clearly resolved (see Fig. 4). These measurements suggest that also at room temperature the cluster phase coexists with the reconstructed Mn islands, however the clusters are mobile and move around the island-free area. Because STM is a very slow method, it cannot image the moving clusters, or only pick up some signal when a cluster is passing under the tip; likely also the tip initiates cluster movement, which could explain the agglomeration of clusters seen at $T$ = 110 K (see Supplementary Fig. S6). This scenario is supported by the observation that while at $T$ = 4 K all clusters are stationary, at $T$ = 80 K a hopping of atoms within the low dilute clusters (DT') is observed (Figure 4 and Supplementary Fig. S4). This makes it plausible that at higher temperature all trimers are mobile, supporting the scenario of supersaturation and step-wise transfer of many Mn atoms from the cluster phase to the island phase.

The occurrence of the cluster phase, and its existence in a metal-on-metal system, is one of the most intriguing results of this study. At very low coverage, on the order of 0.5% of a monolayer, we observe a dilute cluster phase with the same trimer configurations (see Supplementary Fig. S7). We also observe a few dimers or single Mn atoms. Rarely also larger triangular compounds are found, which likely consist of three trimers; possibly they are formed at defects and constitute nucleation points for subsequent island growth (see Supplementary Fig. S8). However, the trimers, and in particular the high compact clusters, dominate. This observation shows that already in the initial stages of growth the formation of trimers is energetically favorable, but the addition of further atoms to such a cluster does not occur. Consequently, there must be an attractive interaction between monomers and also between dimers and monomers, as is typical for metal-on-metal growth. In contrast, the interaction between trimers and monomers must be repulsive.

One tentative explanation for a repulsive interaction between trimers and monomers can originate from strain, when the preferred Mn-Mn bond length is longer than the Ir-Ir distance. Such a large Mn-Mn distance is realized in the expanded reconstructed Mn island phase. If the Mn atoms prefer a relatively long bond distance also in the trimers, then the atoms move outwards from the perfect hollow sites. This would make it more difficult for additional Mn atoms to attach, because it would necessarily mean that they cannot also adsorb near a hollow site but must reside in more unfavorable adsorption sites. This strain related effect might be responsible for the formation of isolated trimers in the cluster phase. We would like to point out that there is no surface state for Ir(111) that crosses the Fermi energy [27], which could lead to a modulation of the diffusion potential due to standing electron waves as seen for noble metal (111) surfaces [5], [6].

Another possible source for a repulsive interaction between clusters is an electric dipole moment. Clusters with dipole moments normal to the surface would naturally repel each other. The strong asymmetry for the switching (Figure 5) suggests that the Mn clusters could indeed exhibit such an electric dipole. Also in the system of Eu on graphene on Ir(111) such a dipole moment was found to be the origin for the formation of a cluster phase, as deduced from DFT calculations [10]. Furthermore, rare earth elements on W(110) have shown a coverage-dependent morphology, where chains with decreasing distance are observed for increasing coverage, which was also explained by electric dipole moments of the rare earth atoms [26]. As we do not have corresponding calculations for our system, the presence of an electric dipole moment of Mn trimers on Ir(111) remains elusive, but it could be a candidate to explain the formation of the cluster phase.

One interesting question regards the magnetic state of the Mn trimers. In SP-STM experiments we were not able to detect a magnetic signal from the clusters. Based on our result of an antiferromagnetic Néel ground state for both the reconstructed island phase and the pseudomorphic phase, it is natural to assume that the Mn atoms also prefer an antiferromagnetic alignment in the trimers. This would again lead to 120° between all three magnetic moments, and thus a compensated total moment for each cluster. Note that we can exclude that this magnetic state is involved in the repulsion between the trimers, because at the temperature where the clusters are mobile, they are expected to be well above the blocking temperature, i.e. in a (super)paramagnetic state. We speculate that at $T$ = 4 K the Mn clusters are antiferromagnetic with 120° between all atoms, but that fluctuations of this magnetic state, i.e., coherent rotations of the coupled spins around the surface normal, inhibit an imaging with SP-STM.

In conclusion we have demonstrated the self-assembly of a monodisperse cluster phase in a metal-on-metal system. We have presented different possible mechanisms for this unexpected observation.

**Methods:**

The experiments were performed in multi-chamber ultra-high vacuum systems. Ir(111) was cleaned by cycles of Ar-ion sputtering with annealing to about 1600°C to recover atomically flat terraces. Occasionally oxygen annealing with temperature ramps up to about 1600°C in partial pressures of oxygen of about $1*10^{-7}$ mbar was performed to remove carbon. Mn was evaporated from a Knudsen cell at 690°C with a deposition rate of about 0.1 atomic layers per minute. The samples studied at T > 100 K were prepared with Mn deposition from an e-beam evaporator. The time delay between the last anneal of the Ir(111) crystal and the Mn deposition is a measure for the substrate temperature during Mn growth. The samples were transferred in-vacuo into the STM. Note that three different home-built STMs were used which are suitable to study the sample in the different temperature regimes.

**Acknowledgements:**

This project has received funding from the European Union's Horizon 2020 research and innovation programme under the Marie Skłodowska-Curie grant agreement No 955671 (SPEAR), and from the Deutsche Forschungsgemeinschaft (DFG, German Research Foundation) via project no. 402843438. We gratefully acknowledge technical support from F. Zahner.

**References:**

[1] Henry C.R., 2D-Arrays of Nanoparticles as Model Catalysts. *Catalysis Letters*, **2015**, 145, 731-749.

[2] Jena P.; Sun Q., Super Atomic Clusters: Design Rules and Potential for Building Blocks of Materials. *Chem. Rev.* **2018**, *118*, 5755–5870.

[3] Brune H., Microscopic view of epitaxial metal growth: nucleation and aggregation, *Surf. Sci. Rep*. **1998**, *31*, 121-229.

[4] Einax M.; Dieterich W.; Maass P., Colloquium: Cluster growth on surfaces: Densities, size distributions, and morphologies. *RMP*. **2013**, *85*, 921.

[5] Silly F.; Pivetta M.; Ternes M.; Patthey F.; Pelz J. P.; Schneider W. D., Coverage-dependent self-organization: from individual adatoms to adatom superlattices, *New J. Phys*. **2004**, *6* ,16.

[6] Cao R.X.; Zhang X.P.; Miao B.F.; Zhong Z.F.; Sun L.; You B.; Hu A.; Ding H.F., Self-organized Gd atomic superlattice on Ag(111): Scanning tunneling microscopy and kinetic Monte Carlo simulations. *Surf. Sci.* **2013**, *610*, 65–69.

[7] Becker C.; Rosenhahn A.; Wiltner A.; von Bergmann K.; Schneider J.; Pervan P.; Milun M.; Kralj M.; Wandelt K., Al2O3-films on Ni3Al(111): a template for nanostructured cluster growth. *New J. Phys*. **2002**, *4*, 75.

[8] Pivetta M.; Rusponi S.; Brune H., Direct capture and electrostatic repulsion in the self-assembly of rare-earth atom superlattices on graphene, *Phys. Rev. B*. **2018**, *98*, 115417.

[9] Will M.; Atodiresei N.; Caciuc V.; Valerius P.; Herbig C.; Michely T., A Monolayer of Hexagonal Boron Nitride on Ir(111) as a Template for Cluster Superlattices. *ACS Nano*. **2018**, *12*, 6871–6880.

[10] Förster D. F.; Wehling T. O.; Schumacher S.; Rosch A.; Michely T., Phase coexistence of clusters and islands: europium on graphene, *New J. Phys.* **2012**, 14, 023022.

[11] Jamneala T.; Madhavan V.; Crommie M. F.; Kondo Response of a Single Antiferromagnetic Chromium Trimer. *Phys. Rev. Lett.* **2001**, *87*, 256804.

[12] Kliewer J.; Berndt R.; Minár J.; Ebert H., Scanning tunnelling microscopy and electronic structure of Mn clusters on Ag(111), *Appl. Phys. A* **2006**, *82*, 63–66.

[13] Hermenau J.; Ibañez-Azpiroz J.; Hübner C.; Sonntag A.; Baxevanis B.; Ton K. T.; Steinbrecher M.; Khajetoorians A. A.; dos Santos Dias M.; Blügel S.; Wiesendanger R.; Lounis S.; Wiebe J.; A gateway towards non-collinear spin processing using three-atom magnets with strong substrate coupling, *Nat. Commun.* **2017**, *8*, 642

[14] Hernández-Vázquez E. E.; López-Moreno S.; Munoz F.; Ricardo-Chavez J. L.; Morán-López J. L., First-principles study of Mn3 adsorbed on Au(111) and Cu(111) surfaces. *RSC Adv.* **2021**, *11*, 31073.

[15] Kurz P.; Bihlmayer G.; Blügel S., Noncollinear magnetism of Cr and Mn monolayers on Cu(111). *J. Appl. Phys*. **2000**, *87*, 6101–6103.

[16] Gao C. L.; Wulfhekel W.; Kirschner J., Revealing the 120 Antiferromagnetic Néel Structure in Real Space: One Monolayer Mn on Ag(111). *Phys. Rev. Lett*. **2008**, *101*, 267205.

[17] Lounis S., Non-collinear magnetism induced by frustration in transition-metal nanostructures deposited on surfaces. *J. Phys.: Condens. Matter*. **2014**, 26, 273201

[18] Brinker S.; dos Santos Dias M.; Lounis S., Prospecting chiral multisite interactions in prototypical magnetic systems, *Phys. Rev. Res.* **2020**, *2*, 033240.

[19] Spethmann J.; M. Sebastian; von Bergmann K.; Wiesendanger R.; Heinze S.; Kubetzka A., Discovery of Magnetic Single- and Triple-q States in Mn/Re(0001). *Phys. Rev. Lett*. **2020**, *124*, 227203

[20] Bode M., Spin-polarized scanning tunnelling microscopy. *Rep. Prog. Phys.* **2003**, *66*, 523.

[21] Wiesendanger R., Spin mapping at the nanoscale and atomic scale. *RMP*. **2009**, *81*, 1495.

[22] von Bergmann K.; Kubetzka A.; Pietzsch O.; Wiesendanger R., Interface-induced chiral domain walls, spin spirals and skyrmions revealed by spin-polarized scanning tunneling microscopy. *J. Phys.: Condens. Matter*. **2014**, *26*, 394002.

[23] Wortmann D.; Heinze S.; Kurz P.; Bihlmayer G.; Blügel S., Resolving Complex Atomic-Scale Spin Structures by Spin-Polarized Scanning Tunneling Microscopy. *Phys. Rev. Lett*. **2001**, *86*, 4132.

[24] Lundgren E.; Leonardelli G.; Schmid M.; Varga P., A misfit structure in the Co/Pt(111) system studied by scanning tunnelling microscopy and embedded atom method calculations. *Surf. Sci*. **2002**, *498*, 257-265.

[25] Heinze S., Simulation of spin-polarized scanning tunneling microscopy images of nanoscale non-collinear magnetic structures. *Appl. Phys. A*. **2006**, *85*, 407-414.

[26] Pascal R.; Zarnitz C.; Bode M.; Wiesendanger R., Atomic and local electronic structure of Gd thin films studied by STM and STS. *Phys. Rev. B.* **1997**, *56*, 3636.

[27] Varykhalov A.; Marchenko D.; Scholz M. R.; Rienks E. D. L.; Kim T. K.; Bihlmayer G.; Sánchez-Barriga J.; Rader O., Ir(111) Surface State with Giant Rashba Splitting Persists under Graphene in Air. *Phys. Rev. Lett*. **2012**, *108*, 066804.